\documentstyle[a4,cite,12pt]{article} 

{\catcode `\@=11 \global\let\AddToReset=\@addtoreset}
\AddToReset{equation}{section}

\newtheorem{Theorem} {Theorem} [section]

\newcommand{\twomanif}{S}
\newcommand{\be}{\begin{equation}}
\newcommand{\ee}{\end{equation}}
\newcommand{\eq}[1]{(\ref{#1})}

\newcommand{\aM}{{\bar M}}              
\newcommand{\aR}{{\bar R}}              
\newcommand{\ame}{{\bar g}}             
\newcommand{\anabla}{{\bar \nabla}}     
\newcommand{\tr}{\mbox{\rm tr}}             
\newcommand{\Weyl}[1]{{W^{(#1)}}}       
\newcommand{\Energy}{\mathcal E}        
\newcommand{\Slice}{{\mathcal S}}       

\renewcommand{\phi}{\varphi}

\title{GR15 Workshop A3 \protect\\  Mathematical Studies of Field
  Equations \\ A report}  
\author{Piotr T.\ Chru\'sciel \\
D\'epartement de  Math\'ematiques, 
Facult\'e des Sciences, \protect\\  
Parc de Grandmont, F--37200 Tours, France}
\begin{document}

\maketitle

\section{Introduction}
In this report I have tried to describe the main streams of research
presented at the workshop. Occasionally I have described some
contributions which were presented in poster form, whenever they were
directly related to the subjects presented in the talks.  This paper
is, essentially, a patchwork of various people description of their
work, which I have edited in an attempt to obtain some uniformity in
style, and to produce a report of finite length. Much to my regret, I
had to cut or shorten a lot of material because of length constraints.
I  wish to thank all those who have sent me a contribution.

\section{The Riemannian Penrose Inequality}

In 1972, Penrose \cite{Penrose:inequality} gave a heuristic argument
to establish the geometric inequality
\be
m_{ADM}(M)\ge\sqrt{|N|/16\pi}
\label{(P)}
\ee for a time-reversible initial data set $(M,g)$, where $|N|$ is the
area of a marginally trapped surface bounding $M$. We refer to
\eq{(P)} as the {\em Riemannian Penrose inequality}, because
throughout this section it is assumed that the initial data have {\em
  vanishing extrinsic curvature}, $K_{ij}=0$. Penrose's argument
relies on knowledge of the future evolution of $M$, including the
Hawking area monotonicity of the (suitably differentiable) event
horizon.  Therefore, Penrose suggested that \eq{(P)} can be seen as a
Riemannian test for the standard picture of black hole formation,
particularly the Weak Cosmic Censorship Hypothesis.  Recently Huisken
and Ilmanen have managed to establish an appropriately understood
version of the inequality \eq{(P)} \cite{HI1,HI2}. This is without
doubt one of the major achievements in mathematical general relativity
in recent years, and the method of proof deserves a short description.
Huisken and Ilmanen's starting point is the argument of Geroch
\cite{Geroch:extraction} and Jang--Wald \cite{JangWald}, which
proceeds as follows: Let $H$ denote the mean curvature and $\nu$ the
outward unit normal of a 2-surface $N$ as a submanifold of $M$.  Let
$(N_t)_{t\ge0}$ be a family of surfaces in $M$ evolving outward with
speed $1/H$, that is, \be {\partial x\over \partial t}={\nu(x)\over
  H(x)},\qquad x\in N_t,\qquad t\ge0,
\label{(star)}
\ee
with $N_0=N$. Then Geroch shows that as long as things go smoothly,
the Hawking mass
$$
m_H(N_t):=
{1\over64\pi^{3/2}}|N_t|^{1/2}\left(16\pi-\int_{N_t}H^2\right)
$$ is monotone nondecreasing during the flow. If $N_t$ becomes a
large, round sphere in the limit, then $m_H(N_t)\to m_{ADM}(M)$.  This
calculation proves \eq{(P)}, provided the surfaces remain smooth.  Now
one does not expect these surfaces to remain smooth in general, and
Huisken and Ilmanen manage to handle that difficulty. A key idea is to
require that each $N_t$ minimizes area among all surfaces surrounding
the ``past history'' $\cup_{s<t}N_s$.  The effect is that $N_t$ must
sometimes jump in order to maintain this condition. This is
implemented through a minimization principle for a function $u$ whose
level sets form the flow. An approximation scheme is used, and
convergence is proven using methods of geometric measure theory.
The final theorem is stated as follows:

\begin{Theorem}[Huisken \& Ilmanen \cite{HI1,HI2}]
 Assume that $M$ is a complete, connected $3$-manifold
which has nonnegative scalar curvature and which is asymptotically flat:
$$
|g_{ij}-\delta_{ij}|\le{C/|x|},\qquad|g_{ij,k}|\le{C/|x|^2},
\qquad R_{ij}\ge-{Cg_{ij}/|x|^2}\ ,
$$
for some flat metric $\delta_{ij}$ near infinity. Suppose that
 $\partial M$ is a compact minimal surface and that $M$ contains no
other compact minimal surfaces.
Then \eq{(P)} holds for each connected component $N$ of $\partial M$.
Moreover equality holds if and only if
$M$ is one-half of the spatial Schwarzschild manifold.
\end{Theorem}

As mentioned previously, this theorem assumes that the initial data
set has vanishing extrinsic curvature\footnote{Actually for all
  purposes of this section it would suffice to assume that the initial
  data surface is maximal and that the extrinsic curvature vanishes at
  the minimal surface.}, so that the general case remains  open.

A result which comes very close to that of Huisken and Ilmanen has
been very recently obtained by Bray: In
\cite{Bray:thesis,Bray:preparation} Bray considers the case where
$(M^3,g)$ has only one outermost minimal sphere $\Sigma_0$. He assumes
that for each volume $V>0$, if one or more area minimizers exist for
$V$, then at least one of these area minimizers for the volume $V$ has
exactly one component. Under this condition he shows that \eq{(P)}
holds.

Another noteworthy new result is that of Herzlich, who establishes an
inequality, somewhat similar to \eq{(P)}, which relates the mass, the
area, and a function theoretic quantity $\sigma$:

\begin{Theorem}[Herzlich \cite{mh-inegalite-penrose}]
Let $(M,g)$ be a $3$-dimensional asymptotically flat Riemannian manifold 
with a compact, connected, (inner) boundary $\partial M$ that is a minimal
(topological) $2$-sphere. Suppose also that the scalar
curvature of $(M,g)$ is nonnegative. Then its mass $m$, if defined,
satisfies
\[
m \geq \frac{1}{2}\,\frac{\sigma}{1+\sigma}\ 
\sqrt{\frac{\mbox{Area}(\partial M)}{\pi}}
\]
where $ \sigma$ is a dimensionless quantity defined as
\[
\sigma = \sqrt{\frac{\mbox{Area}(\partial M)}{\pi}} \ \
\inf_{f\in C_{c}^{\infty}, f\not\equiv 0}
\frac{||df||^{2}_{L^{2}(M)}}{||f||^{2}_{L^{2}(\partial M)}}\ \cdot
\]
Moreover, equality is achieved if and only if $(M,g)$ is a spacelike
Schwarzschild metric of mass $\frac{1}{4}\sqrt{\mbox{Area}(\partial M)/\pi}$.
\end{Theorem}

It would be of interest to study the properties of $\sigma$, in
particular to find out whether $\sigma$ can be large. It has been
pointed out by P.~Tod (private communication) that $\sigma$ tends to
zero for Reissner--Norstr\"om black holes when the extreme limit is
approached. 

The above results relate the ADM mass to the area of a single
component of the minimal surface. When the outermost minimal surface
is not connected one can expect that the inequality can be improved. A
naive way would be to write
\[ m  \ge  \sum_i\sqrt{A_i / 16\pi} \ .\]
We note that this inequality becomes an equality for
Majumdar--Papapetrou black holes.  H.~Bray conjectures (private
communication) that this is the correct generalization;  in his
thesis he proves the following, weaker inequality, under a restrictive
condition: \be
\label{brayineq}
m \ge \left(\sum \left(\frac{A_i}{16\pi}\right)^\frac32 \right)
^\frac13\ .  \ee In order to establish \eq{brayineq}, Bray introduces
the following functional:
\[ F(V) = \inf_{\{\Sigma_i\}} \{ \sum_{i} \mbox{Area}(\Sigma_i)^\frac32 
\,\,|\,\, \{\Sigma_i\} \mbox{ contain a volume $V$ outside the
  horizons} \} \] where the $\{\Sigma_i\}$ are the boundaries of the
components of some 3-dimensional open region in $M^3$. Here one
assumes that and $\bigcup_i \Sigma_i$ is in the closure $\tilde{M}^3$
of that component of $M^3 - \Sigma_0$ which contains the
asymptotically flat end. Moreover in the definition of $F$ one
requires that $\bigcup_i \Sigma_i$ is in the homology class of
$\tilde{M}^3$ which contains both a large sphere at infinity and the
union of the horizons. Bray's condition reads as follows: For each
$V>0$, if one or more sets of surfaces minimize $F$ for the volume
$V$, then at least one of these sets
is pairwise disjoint, that is,
$\Sigma_i \cap \Sigma_j = \emptyset$ for all $i \ne j$.

\begin{Theorem}[Bray \cite{Bray:thesis,Bray:preparation}]\label{R6}
Suppose $(M^3,g)$ is complete, 
has nonnegative scalar curvature, contains outermost
minimal spheres with surface areas $\{A_i\}$, is asymptotically flat with 
total mass $m$, and satisfies the above described condition. 
Then \eq{brayineq} holds.
\end{Theorem}

\section{Initial data}

A standard approach to constructing initial data for the Einstein
equations is the so--called conformal method \cite{IsenbergLiving}.
Except for some special situations with symmetries, the only
alternative (more or less) systematic method of obtaining those data
seems to be via the {\em thin sandwich problem}. This problem is known
to be ill posed in general \cite{ChristodoulouFrancaviglia}, but is
known to be solvable under some conditions \cite{BartnikFodor}. One
advantage thereof is that the restriction of constant extrinsic
curvature, which is standard in the conformal approach, does not
arise. Another is that one has complete control of the initial data
metric. In his contribution to this problem D.~Giulini considers
Einstein gravity on $\Sigma\times R$, where $\Sigma$ is a compact
Riemannian manifold, coupled to a gauge field (with compact gauge
group $G$ and trivial bundle), and $n$ real scalar fields carrying
some representation of $G$. It is assumed that there are no couplings
to 2nd (or higher) derivatives of the metric in the matter Lagrangian.
Let $\Phi_{A}$ collectively denote all fields.
In this setup the thin sandwich problem consists of prescribing
$\Psi:=\{\Phi_A,\dot{\Phi}_A\}$ and trying to solve the scalar
constraint, the vector constraints, and the Gauss constraints for the
lapse $\alpha$, the shift $\beta$, and the Lie--algebra valued gauge
functions $\lambda$.  Giulini assumes that
I.) the generalized De Witt metric $G^{AB}$ satisfies
$G^{AB}\gamma_A\gamma_B<0$, and II.) the potential energy of all
fields $U$ is strictly positive. This allows one to algebraically
solve for $\alpha$ and leads to a system $F(X,\Psi)=0$ of non-linear
PDE's for $X:=(\beta,\lambda)$, with $F$ --- a smooth map between
appropriate Sobolev spaces. Let $X'$ be a solution for given $\Psi$
which satisfies I and II. Under conditions I.) and II.) Giulini proves
that 1.)  $\partial_XF(X',\Psi)$ is a linear, self-adjoint, elliptic
operator (from the space of $X$'s to itself), and 2.)  has trivial
kernel iff ${\bf S}$: $D_{\beta,\lambda}\Phi_A=\alpha{\dot\Phi}_A$,
the symmetry-equation projected to $\Sigma$, implies
$\alpha=\beta=\lambda=0$. Hence, if ${\bf S}$ has just trivial
solutions, the implicit function theorem implies that $F(X,\Psi')=0$
can be uniquely solved in terms of $X(\Psi')$ for any $\Psi'$ in a
 neighborhood of $\Psi$.  This result generalizes the results of
Bartnik and Fodor \cite{BartnikFodor} to the presence of matter fields
of the kind mentioned above.

The talk by S.~Husa was concerned with application of conformal
compactification and conformal symmetries to the (numerical)
construction and analysis of asymptotically flat (AF) vacuum initial
data within the conformal approach.  Instead of picking an AF metric
as 'input' for the Lichnerowicz equation, Beig and Husa choose a
conformally compactified representative of the conformal equivalence
class.  This leads to a convenient treatment of asymptotic regions,
provides a simple construction of (multi--)black hole initial data,
and allows one to exploit conformal symmetries.  In particular the
$U(1)\times U(1)$ conformal symmetry of the physical metric is used by
Beig and Husa to derive a class of exact solutions to the momentum
constraint, and to decompose the time symmetric conformal factor in a
double Fourier series on the group orbits \cite{RBSH94}.  The
solutions are then given in terms of a countable family of uncoupled
ODE's on the orbit space.  The existence of positive solutions is
obtained by computing the sign of the first eigenvalue of the
conformal Laplacian, which in this case becomes an ODE problem.  The
authors have carried out a numerical analysis (including the existence
and properties of apparent horizons) for (i) Brill waves (plus black
holes), (ii) initial data containing a marginally outer trapped torus
\cite{SH96}, and (iii) time-asymmetric initial data, where the
extrinsic curvature is obtained as an exact solution for
non-conformally flat geometries with conformal symmetry.

\section{The evolution question}
\subsection{Local aspects}\label{local}
For hyperbolic partial differential equations two classes of problems
are usually well posed: the initial value problem, also called the
Cauchy problem, and the initial -- boundary value problem. While for
the Einstein equations the local aspects of the former problem are
well understood (\emph{cf., e.g.,} \cite{RendallLiving}), the first
general result about the latter is a theorem of Friedrich and Nagy
\cite{FriedrichNagy} announced during the workshop. Now there are
various ways of stating the problem, and here we will only describe
one of the results obtained, the reader is referred to
\cite{FriedrichNagy} for more general results.

Consider, then, a space-like hypersurface $\Sigma $ and a time-like
hypersurface $T$ in a vacuum space--time $(M, g)$, which intersect
along a space--like two dimensional surface $\twomanif = \Sigma \cap
T$.  One can then ask the question, which data need to be given on
$\Sigma $ and $T$ so that one can reconstruct (by solving Einstein's
equation) $g$ on an appropriate neighborhood $M'\subset M$ of a given point
$p\in \twomanif$, with $M'$ --- one-sided with respect to $\Sigma $
and $T$ (``1/2 sided''). To answer this question, the authors perform
the following construction: Let $x^3$ be a coordinate on $M'$ such
that $T_c = \{x^3 = c = const. \geq 0\}$ is time-like, with $T = T_0$.
Let, next, $e_k$ be a smooth orthonormal frame field on $M'$ with the
property that the time-like vector field $e_0$ is orthogonal to
$\Sigma \cap T_c$ and satisfies $e_0(x^3) = 0$. The field $e_3$ is
chosen to be orthogonal to $T_c$, while the fields $e_A$, $A = 1, 2$,
are tangent to $\Sigma \cap T_c$ and Fermi propagated in the direction of
$e_0$ with respect to the intrinsic connection $D$ induced on $T_c$.
The coordinates $x^0$, $x^1$, $x^2$ are chosen so that $\{x^0 = 0\} =
\Sigma$, $e_0(x^{\mu}) = \delta^{\mu}\,_0$.  Let $\chi(x^0, x^1, x^2, c)$
denote the trace of the second fundamental form of $T_c$. On $T$ one
finally defines
\[C_{AB} = C_{\mu\nu\lambda\rho} e_0^{\mu} e_A^{\nu}
e_0^{\lambda} e_B^{\rho}, \quad c_{AB} =
C_{AB} - \frac{1}{2}\delta_{AB} \delta^{CD} C_{CD}\ ,\]
where $C_{\mu\nu\lambda\rho}$ is the Weyl tensor. Then one of the
results proved in \cite{FriedrichNagy} reads as follows:

\begin{Theorem}[Friedrich, Nagy, 1997]
  \label{TFN}
Let $(h_{\alpha \beta},k_{\alpha \beta})$ be the standard Cauchy data
induced on $\Sigma $ by $g$, let $\chi(x^0, x^1, x^2, 0)$ be the mean
extrinsic curvature of $T$ and let $c_{AB}$ be related to  the
Weyl tensor as described above. Let further $\Gamma_A(x^{\mu}) = g(e_A,
D_{e_0}\,e_0)$ and $\chi(x^{\mu})$, $x^3 > 0$ be some given ``gauge
source functions''.  Then for any point $p\in\twomanif$ there exists a ``1/2
sided'' neighborhood $M'$ of $p$ on which the data determine a
unique, smooth solution to Einstein's field equations. In that
solution the data assume the geometrical interpretation given above.
\end{Theorem}

We emphasize that the result above has been presented as a {\em
  reconstruction problem}. Friedrich and Nagy also consider the
question, how to construct space--times ``from scratch'': in that case
standard {\em corner conditions\/} arise \cite{FriedrichNagy}.

In the workshop T.~Velden gave a talk in which she points out a gap in
the analysis of the Cauchy problem for Einstein equations with
non-rotating dust as a source. She emphasized that the question of
propagation of constraints was not properly handled previously, and
presented an analysis where this problem is taken care of. She also
described a local existence theorem of the Cauchy problem assuming
analyticity of the data. It is likely that the analyticity hypothesis
can be gotten rid of by extracting a symmetric hyperbolic system out
of the system of equations at hand. This can most probably be done by
repeating the analysis carried out by H.~Friedrich in a related
context \cite{Friedrich:fluid}. In that last paper H.~Friedrich
extracts a well posed system of equations out of the Einstein
equations with ideal fluid as a source, {\em without} assuming that
the density of the fluid is bounded away from zero. This is particularly
remarkable, as no such procedure is known in general for the Euler
equations considered by themselves, {\em i.e.\/} when not coupled to
general relativity.

To close this section, let us note that there has been quite a lot of
activity in the last few years concerning the question, how to extract
a well posed system of dynamical equations out of the Einstein
equations
\cite{FGR,AACY,Friedrich:hyperbolicreview,BonaMassoSeidelStela,ReulaLiving}. In
particular A.~Anderson has submitted an abstract to this workshop
describing his construction, in collaboration with Y.~Choquet--Bruhat
and J.~York, of a new first order symmetric hyperbolic system for the
evolution part of the Cauchy problem of general relativity \cite{CBYA}.

\subsection{A semi--global result}

There are only very few global or semi--global existence results
concerning solutions of the Cauchy problem in general relativity (see
\cite{RendallLiving} for an exhaustive list). So far the only ones
which concerned space--times without any symmetries were the 
Christodoulou--Klainerman theorem of stability of Minkowski
space--time \cite{Ch-Kl}, the semi--global stability results of
Friedrich (\emph{cf., e.g.}\ \cite{Friedrich:aDS} and references
therein), and the semi--global existence results concerning the
Robinson--Trautman space--times \cite{ChRT}. This short list has been
recently extended by Andersson and Moncrief with a semi--global
existence result concerning the following class of spatially compact
space--times: Let $M$ be a compact 3--manifold of hyperbolic type and
let $g_0$ be the standard hyperbolic metric on $M$ with sectional
curvature $-1$.  The couple $(M,g_0)$ is called {\em rigid} if there
are no trace free, divergence free 
symmetric 2--tensors $u_{ij}$ satisfying $u_i^{\ i} = 0$, $\nabla^j
u_{ji} =0$, $\nabla_k u_{ij} - \nabla_j u_{ik} = 0$.  The class of
rigid hyperbolic 3--manifolds is nonempty.

On $\aM = M \times \{t>0\}$ define the metric $\ame_0 = - dt^2 + t^2
g_0$. Then $(\aM, \ame_0)$ is a flat globally hyperbolic spacetime
(its universal cover is the $\kappa=-1$ Friedmann -- Robertson --
Walker vacuum spacetime). Given a spacelike hypersurface $M$ of a
spacetime $(\aM,\ame)$, let $g_{ij}, T^a, k_{ij} = - \anabla_i T_j$ be
the metric, future normal and second fundamental form on $M$, where
$\anabla$ is the covariant derivative on $\aM$. With these
conventions, the induced initial data for the Einstein equations on $M
= \{ t=1\}$ in the standard spacetime $(\aM,\ame_0)$ are $(g_0,
-g_0)$.  Rigidity of $(M,g_0)$ is equivalent to rigidity of $(\aM,
\ame_0)$ in the moduli space of flat spacetime structures.  (Note that
Mostow rigidity does not apply in the case of flat spacetime
structures.)

Andersson and Moncrief consider the 3+1 vacuum Einstein evolution
equations, with constant mean curvature time gauge $\tr k = \tau$ and
spatial gauge fixing given by the affine conformal slice (see
\cite{fischer:moncrief:hamred}) $\Slice_{\tau}$.  Note that with the
orientation used, the mean curvature $\tau = \tr_g k$ satisfies $\tau
< 0$ and $\tau \nearrow 0$ corresponds to infinite expansion (and
proper time $t \nearrow \infty$).
\begin{Theorem}[Andersson \& Moncrief \cite{AndMon:preparation}] 
\label{thm:glob}
  Let $M$ be rigid, and let $(g,k) \in \Slice_{-3}$ be vacuum data for
  Einstein equations, sufficiently close in $H^4 \times H^3$ to
  standard data $(g_0, -g_0)$.  Then (1) Global existence in the
  expanding direction, $\tau \nearrow 0$, holds for the 3+1 vacuum
  Einstein evolution equations, gauge fixed as above, with initial
  data $(g,k)$. (2) The maximal globally hyperbolic vacuum extension
  $(\aM, \ame)$ is geodesically complete in the expanding direction.
\end{Theorem}
In contrast to the asymptotically flat spacetimes considered by
Christodoulou--Klainerman \cite{Ch-Kl}, the maximal globally
hyperbolic vacuum developments of data considered by Andersson and
Moncrief are not causally geodesically complete to the past. This
follows from the singularity theorems of Hawking and Penrose. No other
properties of the space--time in the contracting direction are known
(existence of curvature singularities? Cauchy horizons?).

The proof of Theorem \ref{thm:glob} makes use of a Bel--Robinson type
energy function $\Energy$ defined with respect to the the Weyl fields
$\Weyl{0}_{abcd} = \aR_{abcd}$, where $\aR$ is the Riemann tensor of
$(\aM, \ame)$, $\Weyl{1}_{abcd} = T^f \anabla_f W_{abcd}$ and
$\Weyl{2}_{abcd} = T^f T^g \anabla_f \anabla_g W_{abcd}$. See
\cite{Ch-Kl} for background and a different approach to defining
higher order Bel--Robinson energies.  The use of an energy function to
control the gauge fixed evolution used in the proof of Theorem
\ref{thm:glob} is closely related to the method used to prove global
existence in CMC time for 2+1 gravity
\cite{andersson:moncrief:tromba:2+1}.

The Andersson--Moncrief theorem shows once again that the
Bel--Robinson tensor is an object which deserves attention --- recall
that this tensor has already been used by Christodoulou and Klainerman
in their stability theorem \cite{Ch-Kl}. During the workshop
J.~Senovilla has presented a simple proof, obtained using the
Bel--Robinson tensor, of the following fact \cite{BS2}: {\it For any
  conformally vacuum spacetime (with or without cosmological
  constant), if the Weyl tensor vanishes on any closed achronal set
  $\Sigma$, then it vanishes on its domain of dependence ${\cal
    D}(\Sigma)$.} While this result is well known, at least in vacuum,
the previous proofs made use of rather more involved considerations.

\section{The nature of  singularities}

\subsection{Numerical experiments}
\label{numerical}
One of the main challenges of mathematical general relativity is the
description of the generic singularities that arise during evolution
via the Einstein equations out of appropriately regular initial data.
This problem seems to be completely out of reach of the analytical
tools which are currently at our disposal, and our best hope today to
get some real insight about that question is to carry out reliable
numerical experiments. For vacuum Einstein equations two types of
behavior have been known to occur, when a singular boundary (whatever
this means) of a spatially compact space--time is approached: a
``velocity dominated'' behavior (AVTD) and a ``Mixmaster'' behavior.
The latter case is expected to correspond to curvature singularities.
The AVTD behaviour includes both curvature singularities, Cauchy
horizons, and ``topological singularities'' --- those arise when
``what would have been a Cauchy horizon'' is quotiented--out by an
ergodic action of a isometry group \cite{Mess,IKSK}, and leads to
space--times which are inextendible with the curvature remaining
bounded.  A new kind of behavior has been recently observed by
Breitenlohner -- Lavrelashvili -- Maison \cite{lav-BLM} and,
independently by Gal'tsov -- Donets -- Zotov
\cite{lav-DGZ1}\protect\footnote{While D.~Gal'tsov reported on his
  studies in the workshop, G.~Lavrelashvili was kind to contribute to
  this report a description of his results with Breitenlohner and
  Maison. I wish to thank him for this, as well as for many clarifying
  comments concerning his work.}.  While the studies of those authors
are in principle concerned with black hole space--times, one can
spatially compactify the space--time (``below the event horizon'') to
obtain models with $S^2\times U(1)$ topology and $SO(3)\times U(1)$
isometry group.  The numerical analysis of the interior geometry of
static, spherically symmetric black holes of the
Einstein-Yang-Mills-Higgs theory shows the following
\cite{lav-DGZ1,lav-BLM}:

First, within the set
of initial data considered, generically no inner (Cauchy) horizon is
formed inside the non-Abelian black holes. This is then consistent
with the (strong) cosmic censorship hypothesis.

Next, the generic black hole solution of the EYM theory has an
oscillatory behavior inside the horizon~\cite{lav-DGZ1,lav-BLM}.  As
one performs numerical integration starting at the horizon and
integrates towards $r=0$ one observes a sudden steep rise of a
derivative of the {\sl SU(2)\/} gauge field amplitude, $W'$, and a
subsequent exponential growth of the mass function $m(r)$
(parametrizing the $g^{rr}$-component of the metric via
$g^{rr}=1-2m(r)/r$).  Within a short interval of $r$ the mass function
reaches a plateau and stays constant in some range of $r$'s until it
starts to decrease again.  When the solution comes close to an inner
horizon, the same inflationary process repeats itself with an even
more violent next ``explosion''. In the black hole context this
behavior seems to be related to the ``mass inflation'' phenomenon
observed for linear perturbations of the Reissner Nordstr\"om black
holes \cite{BDIM}. In fact, the results of \cite{lav-DGZ1,lav-BLM} can
be thought of as giving a non--linear counterpart of that effect. It
should, however, be emphasized that the dynamics here is essentially
different, because in the linearized case considered in \cite{BDIM} no
plateaux occur. An apparently related ``oscillatory mass inflation''
has also been observed in \cite{Page:inflation}, in a
semi--phenomenological model with two null fluids.
 
By a suitable fine tuning of the initial data at the horizon 
it is possible to obtain solutions with a different (non--generic)
behavior, see \cite{lav-DGZ1,lav-BLM} for details. 
In ~\cite{lav-DGZ1,lav-BLM} one can also find a simplified dynamical
system which seems to provide a qualitative understanding of the
behavior of the generic solutions. In particular one can derive a
``plateau -- to -- plateau formula'' \cite{lav-BLM} which (in the
simplified model) relates quantities at one plateau (before the
``explosion'') to those on the next plateau (after the ``explosion'').
  
In order to study the model dependence of these results the
theory with an additional Higgs field was also
investigated~\cite{lav-BLM}.  It was found that after adding the Higgs
field no more oscillations occur in the asymptotic behavior inside
the horizon.
This change in the behavior of the generic solution can presumably be
understood by a change of the character of fixed points of the
corresponding simplified dynamical system.  In that system the
addition of a Higgs fields makes the focal point disappear, and the
asymptotic behavior becomes governed by a stable attractor
\cite{lav-BLM}.

The main conclusions are, that no inner (Cauchy) horizon are formed
inside non--Abelian black holes in the generic case, instead one
obtains a ``spacelike'' singularity at $r=0$.  Without a Higgs field,
{\em i.e.\/} for the EYM theory, one obtains a kind of ``mass
inflation'' that repeats itself in cycles of ever more violent growth.
This behavior near the singularity does not have a counterpart in the
dynamics of spatially homogeneous vacuum cosmological models. With the
Higgs field no such cycles occur in the asymptotic behavior.
 
The next numerical study which has been reported on in this workshop
concerns space--times with $U(1)$ or $U(1)\times U(1)$ symmetry.
Recall that Grubi\u{s}i\'{c} and Moncrief \cite{grubisic93,grubisic94}
have used formal asymptotics expansions as a tool to understand the
asymptotic approach to the singularity in spatially inhomogeneous
cosmologies. They have found that solutions in the form of formal
series could be obtained with the following property: When approaching
the singularity the formal solutions approach a solution of the
equations obtained by dropping, in Einstein's equations, terms
containing spatial derivatives.  This is the AVTD behaviour alluded to
above. (It has been recently shown by Kichenassamy and Rendall that
the Grubi\u{s}i\'{c}--Moncrief formal series are convergent in some
cases \cite{KichenassamyRendall}.)
Self--consistent formal solutions with an AVTD
singularity \cite{isenberg90} at (say) $\tau = \infty$ are obtained if
the terms neglected in the truncated equations are exponentially
small, when evaluated using the AVTD solution. For example, in vacuum
Bianchi IX (Mixmaster) \cite{belinskii71b,misner69} space--times, the
substitution of the Kasner solution into the minisuperspace potential
always yields exponential {\em growth} in one of the terms, which is
not consistent with an AVTD behavior. On the
other hand, in the (plane-symmetric, vacuum) Gowdy cosmologies on $T^3
\times R$ \cite{gowdy71}, nonlinear terms in the wave equations allow
the (formal) AVTD solution 
provided an ``asymptotic velocity $v$'' satisfies $0 < v(\theta) <
1$. 
Numerical simulations performed by Berger, Garfinkle and Moncrief show
how $v(\theta)$ outside the allowed range is driven into the allowed
range by these same nonlinear terms \cite{berger97b,berger97c}. Gowdy
models with a modified topology generalized to include a magnetic
field become the spatially inhomogeneous generalization of magnetic
Bianchi VI$_0$ homogeneous solutions that are known to display
Mixmaster dynamics \cite{leblanc95,berger96a}. The extra nonlinear
terms from the magnetic field cause AVTD behavior to be inconsistent
for essentially all values of $v(\theta)$. The Mixmaster approach to
the singularity one then expects has been observed numerically
\cite{weaver98}. It should be emphasized that this is a first
numerical observation of Mixmaster behavior in a non--homogeneous
space--time. A further generalization to $U(1)$ symmetric cosmologies
on $T^3 \times R$ \cite{moncrief86,berger93} shows that the AVTD
solution is consistent for polarized models but inconsistent for
generic ones. This leads one to expect a Mixmaster--like singularity
at each spatial point. Numerical simulations provide strong support
for an AVTD singularity in the polarized case \cite{berger97c}. In the
generic case, numerical results are suggestive but numerical errors
associated with the failure to preserve the Hamiltonian constraint
prevent strong conclusions \cite{berger97c,VinceBev}. 

\subsection{Other studies}

The singularity theorems of Hawking and Penrose predict geodesic
incompleteness, and many authors identify this feature with the
existence of a singularity. While this attitude is justified in many
cases, situations occur in which one could question the validity of
this conclusion.  In his talk, C.~Clarke summarized the philosophy of
regarding as singular only those points in the space-time manifold
(with a metric that was not necessarily continuous) at which the
propagation of test-fields was disrupted. More precisely, he proposes
to use uniqueness and existence results for the wave equation in a
neighborhood of a point to classify points as ``regular'' or
``singular''. Such an approach allows one to class as regular some of
those points which would be classed as singular in terms of the
completeness of unique geodesics.  He announced a theorem concerning
space-times
for which 1) $g_{ij}$ and $g^{ij}$ are continuous with square
integrable weak derivatives, and which 2) have a point $p$ such that
$g_{ij}$ is in $C^1(M\backslash J^{+}(p))$. Moreover he requires
that, in a coordinate neighborhood of $p$, the integrals $
I_\gamma(a):=\int_0^a\left\vert\Gamma^i_{jk}(\gamma(s))\right\vert ^2
ds$ and $ J_\gamma(a):= \int^a_0\left\vert
  R^i{}_{jkl}(\gamma(s))\right\vert ds $ are bounded by positive
functions $M(a), N(a)$ tending to zero with $a$, for all curves
$\gamma$ whose tangent vector had components lying in some fixed cone
$C$. Under those conditions, he claimed, the wave equation has unique
solutions, in $H^1(S_t)$ for a slicing by spacelike hypersurfaces
$S_t$, in a neighbourhood of $p$ for $C^2$ data on a partial Cauchy
surface to past of, and sufficiently close to, $p$.

He suggested that this theorem might be applicable to shell crossing
singularities, which would become regular points on this proposed
definition. If this were the case, and if one reformulates the cosmic
censorship question according to his proposal, then such singularities
would stop being counter-examples to cosmic censorship, without the
special pleading of dismissing the dust matter as unphysical. He also
claimed that a similar treatment could be applied to cosmic strings.

Since the pioneering work of Belinski, Lifschitz and Khalatnikov
already mentioned above \cite{belinskii71b}, a tool which has been
often used when trying to understand the nature of singularities is
that of formal expansions (\emph{cf.\/} also
\cite{grubisic93,grubisic94}). On one hand, both 1) the numerical
results of Berger and collaborators described above and 2) the
analytical results of \cite{isenberg90,CIM,KichenassamyRendall} give
strong support to the validity of such an approach in some situations.
On the other hand, it is not clear that the Mixmaster, or the
``plateau to plateau'' behavior of Breitenlohner {\em et al.} ---
Gal'tsov {\em et al.} described in Section \ref{numerical} for the
Einstein--Yang--Mills equations, are compatible with any formal
expansions framework.  Whatever the status of such expansions in
general, it is of interest to find if the evolution equations for some
other gravitating systems are (perhaps formally) compatible with this
idea.  L.~Burko has used such an approach in the context of spherical
charged black holes perturbed nonlinearly by a self-gravitating,
minimally-coupled, massless scalar field. As in the
Einstein--Yang--Mills--Higgs case, this model can be spatially
compactified under the horizon to give a cosmological space--time.
The numerical simulations described in \cite{BradySmith,BurkoPRL}
suggest that in this case the Cauchy horizon turns into a ``null weak
singularity'' which ``focuses'' monotonically to $r=0$ at late times,
where the singularity ``becomes spacelike''\footnote{All the terms in
  inverted commas here have an obvious meaning in the coordinate
  system used. I have used inverted commas because I am not aware of
  any standard meaning of those notions in general.}. Burko
\cite{BurkoIOP,Burko:prepa} examines a formal series--expansion
solution for the metric functions and for the scalar field near $r=0$
under the simplifying assumption of homogeneity. He finds that such
solutions are self--consistent, and {\em generic} in the sense that
the solution depends on the expected number of free parameters. The
properties of the formal solutions are similar to those found in the
fully-nonlinear (and inhomogeneous) numerical simulations.

Cauchy horizons are objects which are closely related to singularities
in many situations, so a study of those complements naturally that of
singularities.  T.~Helliwell and D.~ Konkowski have presented a poster
in which they describe some progress in their program to study the
stability of Cauchy horizons \cite{KonkowskiHelliwell}: In
\cite{KonkowskiHelliwellpreprint} they study vacuum plane wave
spacetimes and assert, among others, that there exists a
relationship between the stability of the Cauchy horizon and the
behaviour of test fields in the space--time, provided there is no
Weyl--tensor singularity in the spacetime.

\section{Global techniques}

In his talk describing joint work with L.~Andersson and R.~Howard, G.\ 
Galloway discussed some results about { cosmological time functions},
defined as follows: Let $(M,g)$ be a time oriented Lorentzian
manifold.  The {\it cosmological time function\/} of $M$ is the
function $\tau: M \to (0,\infty]$ defined by: $\tau(q)=\sup_{p<
  q}d(p,q)$, where $d$ is the Lorentzian distance function.  In the
cosmological setting, $\tau$ has a simple physical interpretation:
$\tau(q)$ is the maximum proper time from $q$ back to the initial
singularity.  In general, $\tau$ need not be well-behaved.  For
example, even if $M$ is globally hyperbolic and $\tau$ is finite
valued, $\tau$ need not be continuous (\cite{AGH,WY}).  Define
$\tau$ to be {\it regular\/} provided it satisfies: $\tau(q)<\infty$
for all $q$ and $\tau\to 0$ along every past inextendible causal
curve. In \cite{AGH} some consequences of regularity are obtained.  It
is shown, for example, that if $\tau$ is regular then (1)~$M$ is
necessarily globally hyperbolic, (2) $\tau$ is a time function in the
usual sense, {\em i.e.}, $\tau$ is continuous and strictly increasing
along future directed causal curves, (3)~$\tau$ is {\it
  semi-convex\/}, and hence has first and second derivatives almost
everwhere and (4)~every point of $M$ can be connected to the initial
singularity by a timelike geodesic ray that realizes the distance to
the singularity. See \cite{AGH} for further results.

It should be mentioned that Wald and Yip \cite{WY} introduced the
cosmological time function (or rather its time dual, which they
referred to as the ``maximum lifetime function") quite some time ago
in order to study the existence of synchronous coordinates in a
neighborhood of a spacelike singularity.  More recently, Andersson and
Howard \cite{AH} have made use of the cosmological time function to
prove some results concerning the rigidity of Robertson-Walker and
related spacetimes.

In his talk, E.~Woolgar reported on work in progress with G.~Galloway
concerning the generalization of the topological censorship theorem
\cite{FriedmanSchleichWitt,Galloway:fitopology} to spacetimes with
more general asymptotic 
structures than the usual asymptotic flatness.  Galloway and Woolgar
\cite{GallowayWoolgar} make no assumptions on the geometry near
infinity, and therefore allow timelike, spacelike, and null scris.  In
particular, they show that $\pi_1$ of the domain of outer
communications is a subgroup of $\pi_1$ of scri.  For this one needs
to assume a null energy condition and the null generic condition,
although there are some reasons to believe that the latter assumption
may be unnecessary. It is also assumed that no compact set of
spacetime contains all of scri within its past, and that there are no
naked singularities (in the sense that there is no incomplete null
geodesic whose past is contained within the past of a point of scri).

We note that black hole horizons with non-zero genus in locally
anti-de Sitter spacetime  have been the subject of much recent
attention \cite{ABBHP,ABHP,HolstPeldan,BrillaDS}. The work by Galloway
and Woolgar sheds light on the compatibility of those solutions with
the notion of topological censorship.

P.~Aichelburg and F.~Schein contributed to the workshop an interesting
new time machine: They have constructed a static axisymmetric wormhole
from the gravitational field of two charged shells which are kept in
equilibrium by their electromagnetic repulsion.  
The interior of the wormhole is a Reissner-Nordstr\"om black hole
matching the two shells.  The wormhole is one-way traversable, {\em
  i.e.}, one throat lies in the (local) future of the other and
connects to the same asymptotic region. Moreover, the shells of matter
can be choosen to satisfy the energy conditions.  The solution is an
``eternal time machine'': every point in the Majumdar-Papapetrou region
lies on a closed timelike curve.

In his talk F.~Stahl, described some results he obtained in his
studies of the Schmidt metric, and made some remarks on his intended
further studies. Recall that the Schmidt metric $G$ is a Riemannian
metric on the frame bundle $LM$ of a manifold $M$ with connection,
related to the generalized affine parameter length of curves in $M$
\cite{Schmidt:b-boundary}. St{a}hl has shown \cite{Stahl} that
geodesics of $G$ project down to geodesics on $M$. He hopes to be able
to find a relationship between the conjugate points in $(LM,G)$ and
$M$, and to explore further the properties of singularities by
studying the structure of $(LM,G)$ near a $b$-boundary point.

\section{Stationary solutions}

There have been many attempts to construct models of stationary stars
using the Einstein equations with a perfect fluid. The results
obtained so far include the famous disk models of Meinel and
Neugebauer (\emph{cf.\/} \cite{Meinel:Neugebauer:Kleinwaechter} and
references therein), and the slowly rotating solutions of Heilig
\cite{Heilig}.  It is, nevertheless, clear that much more research is
needed before we get a good understanding of the problems involved.
In his talk M.~Mars described his work with J.~Senovilla \cite{MS1} in
which one discusses how 
to construct the exterior gravitational field created by a given
(arbitrary) distribution of stationary and axisymmetric matter. The
matching conditions \cite{MS} between the known interior and the
unknown vacuum exterior introduce two new, essential parameters in the
interior metric which are closely related with the state of motion of
the body as seen from infinity.  For each value of these parameters,
the exterior matching boundary is uniquely determined. Furthermore,
the matching conditions provide the boundary conditions for the
exterior problem.  In terms of the Ernst potential ${\cal E} \equiv
e^{2U} + i\, \Omega$ \cite{Ernst}, the values of $U$ and the normal
derivatives of ${\cal E}$ are fixed on the boundary and the values of
$\Omega$ are fixed on the boundary up to an arbitrary, non-trivial,
additive constant. Since the equations to be solved are elliptic 
of second order the
problem is not well-posed, so that not every interior metric can be
interpreted as an isolated rotating body. 
Mars and Senovilla prove uniqueness of the exterior field using, in a
first step, the well-known Mazur--Bunting identity \cite{Bu} for harmonic
maps between manifolds to show that the exterior field is unique for
each value of the undetermined additive constant in $\Omega$ on the
matching hypersurface. The second step uses the high degree of
symmetry of the target manifold of the harmonic map in order to
construct a three-parameter conserved current which is then exploited
to show that only one of the additive constants in $\Omega$ is
possible, thus completing the proof.  Mars also described some related
results by Weinstein which, while mostly concerned with stationary
black holes, also do apply to the problem at hand (see \cite{Weinstein3}
and references therein).

While the results of Mars and Senovilla concerned uniqueness,
U.~Schaudt, in collaboration with H.~Pfister, has proved both
existence and uniqueness in a ``small data'' context. (The
qualification ``small'' here does not refer to some abstract small
number, but to a precise explicit bound given by the authors, which might
actually be considered as large in some physical contexts.)
In \cite{Pfister:Schaudt,Schaudt} it was shown that the Dirichlet
problem for the vacuum region outside a ball, as well as for a ball
inside the matter, has a unique regular solution if the absolute
values of the appropriate boundary data are limited by an
appropriately understood ``radius'' of the ball.  The results obtained
include an existence and uniqueness--proof for the exterior solutions
in a range of parameters which includes  all known white
dwarf--stars. 
The method of proof (which is a fixed point argument in appropriately
chosen spaces) and the results have connections with a numerical
solution technique for rotating stars \cite{BGSM}.  A proof of K.
Thorne's ``hoop--conjecture'' \cite{Thorne:collapse} is also given
under, however, a rather restrictive set of hypotheses.

A completely different approach to the construction of stationary
stars is that using Riemann--Hilbert techniques, discussed by C.~Klein
in his talk. Those techniques provide an important tool in the context
of integrable equations since they can be used to generate solutions
with a free function. If this function can be determined from a
boundary or initial value problem, the Riemann--Hilbert problem (RHP)
is equivalent to the solution of a linear integral equation. In the
case of the Ernst equation, Klein and Richter prove two theorems: 1.
The RHP for the Ernst equation (see \cite{KleinRichter:PRD}) with
analytic `jump data' is gauge equivalent to a scalar problem on a
four--sheeted Riemann surface. For rational jump data, it can be
solved explicitly in terms of hyperelliptic theta functions. 2. The
obtained (normalized) solutions are regular except at a contour in the
meridian plane that corresponds to the contour of the RHP. They are in
general asymptotically flat, have a regular axis of symmetry, and can
have ergoregions and an ultrarelativistic limit, see
\cite{KleinRichter:PRL}. These theorems suggest that it might be
possible to solve boundary value problems for the stationary
axisymmetric vacuum Einstein equations by directly identifying the
branch points of the Riemann surface and the free function of the RHP
from the boundary data.

\section{Distributions and their generalizations}

Singular null hypersurfaces on a spacetime manifold, {\em i.e.\/} null
hypersurfaces across which the metric tensor is only $C^0$, can be
used as models for the propagation of an impulsive lightlike signal
({\em e.g.\/} a burst of neutrinos together with an impulsive
gravitational wave).  Using a formalism developped by Barrab\`es and
Israel \cite{BarrabesIsrael}, C.~Barrab\`es, G.~Bressange and P.~Hogan
have shown that in the general case a lightlike shell coexists with an
impulsive gravitational wave on such an hypersurface. They have
analyzed in detail two examples illustrating this phenomenon: The
first concerns abrupt changes of the multipole moments in the
axisymmetric Weyl solutions, and can be used as a model for a
supernova \cite{BBH1}. The second describes jumps in the mass and
angular momentum in Kerr spacetime, and is supposed to model pulsar
glitches \cite{BBH2}.

While the above work concerned distributions within a well posed
mathematical framework, it is often the case that in general
relativistic considerations one encounters products of distributions,
leading to various problems as far as the mathematical meaning of the
resulting expressions is concerned.  Steinbauer has been studying
those issues in the context of a study of geodesics in space--times with
impulsive pp--waves.  Here the main problem to deal with are the
products ``$\theta\delta$'', ``$\theta^2\delta$'' and ``$\delta^2$'',
which arise due to the presence of the Dirac-$\delta$-``function'' in
the metric.  He regularizes those terms by a natural procedure which
corresponds to the physical idea of viewing the impulsive wave as the
limiting case of a sandwich wave. He shows that in this way one can
end up with regularisation independent distributional solutions,
without recourse to ad-hoc strategies which had  been used
in previous analyses of this problem.

In his talk J.~Wilson has described how Colombeau's theory of
generalised functions \cite{Colombeau}, which gives a mathematical
framework in which products of distributions are well defined objects,
can be used to calculate the curvature of metrics which are too
singular to be handled by the standard distribution theory.  Detailed
calculations have been performed for static cosmic strings in flat
space-time \cite{ClarkeVickersWilson} and for dynamical cosmic strings
on curved backgrounds \cite{Wilson}. In particular it was shown that
the mass per unit length of a radiating cosmic string, whose metric
takes the form
$$ ds^2=e^{2\gamma(t-r)}(-dt^2+dr^2)+r^2d\phi^2+dz^2, $$
is $2\pi(1-e^{-\gamma(t)})$, which agrees with the mass at null
infinity.  In this context it should be pointed out that Colombeau's
theory itself is not invariant under $C^\infty$ diffeomorphisms, so
one must exercise some caution in choosing the coordinate system when
applying it to a covariant physical theory. 
In his talk J.~Wilson expressed the hope to be able to construct such
a diffeomorphically invariant theory of generalised functions, basing
on a recent paper by Colombeau and Meril \cite{ColombeauMeril}.

It should be pointed out that R.~Mansouri and K.~Nozari have presented
a poster describing their work on a distributional approach to the
change of signature of spacetime metric, based again on Colombeau's
generalized functions. They claim to obtain an Einstein equation for
dynamics of signature changing spacetime with a nonvanishing
distributional stress-energy tensor supported at the signature--change
hypersurface.

Let us finally note that V.~Pelykh has contributed to this workshop an
abstract describing an approach to handle Einstein equations with
distributional sources.

\section{The energy of the gravitational field}

A very old problem, which still attracts some attention, is that of
energy of the gravitational field. In his talk J.~Nester reported on
his attempts, with C.C.~Chang and C.M.~Chen, to construct
gravitational Hamiltonians for finite regions. (A very similar
approach has already been advocated by Kijowski, see
\cite{Kijowskiold,KijowskiGRG} and references therein).  Those
Hamiltonians consist of boundary integrals, the numerical value of
which can be thought of being the quasi--local energy contained in the
volume enclosed. The variation of the Hamiltonian leads to surface
terms which determine what needs to be held fixed on the boundary if
one wishes to obtain a Hamiltonian system.  There isn't therefore only
one energy because there are many boundary condition which one might
wish to impose, each with its distinct boundary term.  Nester and his
collaborators examine specific cases, identifying the appropriate
boundary conditions.  They suggest that certain principles (good
limiting values, covariance \cite{CNT}) can be used to restrict the
possibilities. Nester emphasizes that one can give a ``respectable''
quasilocal interpretation to the pseudo--tensors, and claims that
 superpotentials associated with  pseudotensors form  legitimate
Hamiltonian boundary terms. 
It would be of interest to make a detailed Hamiltonian
analysis, along the lines presented by Kijowski in
\cite{KijowskiGRG} and also by Nester, of those boundary conditions
which lead to a well posed Cauchy problem, such as {\em e.g.\/} the
Friedrich--Nagy conditions described in Section \ref{local}.

A different approach was described in a poster by Petrov, who
advocates the use of a background metric to define the
energy--momentum tensor of the gravitational field, and hence the
local and quasi--local energy. In a joint work with Narlikar
\cite{PetrovNarlikar} they assert that one can give a distributional
meaning to this tensor calculated for the Schwarzschild and the
Reissner-Nordstr\"om solutions with respect to an appropriately chosen
background.

Yet another approach to the energy question was described by Yoon, who
proposes a derivation based on the Hamiltonian formalism of general
relativity in the (2,2)--foliation of spacetime. Here one uses {\it
  Kaluza--Klein} type variables as the configuration variables.  The
analysis is carried out in the Newman--Unti gauge, which partially
fixes the spacetime diffeomorphisms. An ``advanced time'' is used as
the time coordinate on which the Hamiltonian formalism is based.  The
four Einstein's constraints are written in the canonical variables,
and are found to be the first--class constraints generating the
residual spacetime diffeomorphisms that survive after the Newman--Unti
gauge is chosen. By integrating one of the constraint equations over a
closed spacelike 2--surface one obtains an integral equation which
relates the rate of change of a certain integral over the 2--surface
to a flux integral across that 2--surface. It is suggested to define
the quasi--local gravitational energy of a region enclosed by the
2--surface via the former integral. The latter integral becomes then
the gravitational energy flux integral crossing the 2--surface.  The
proposed quasi--local integrals reproduce the Bondi energy and Bondi
flux integral at null infinity.

Let us also mention that N.~Dahdich has presented a poster in which he
analyzes the properties of the Brown--York \cite{YorkBrown} quasilocal
energy in spherically symmetric black hole space--times.

It is clear that all those approaches lead to a plethora of
definitions of energy, while one would like to have a unique one. In a
poster by J.~Jezierski, M.~McCallum and this author a uniqueness
theorem for the Trautman--Bondi mass was presented, which can be used
to single out this energy among the competitors \cite{CJMPRL}.

\section{Miscellaneous}

In his talk F.~Andersson presented some new results concerning the
Lanczos potential \cite{Lanczos}, due to himself and to S.B.~Edgar and
A.~H\"oglund. Recall that the original proof of the existence of that
potential by Lanczos was flawed, and the first complete proof was
given by F.~Bampi and G.~Caviglia \cite{BampiCaviglia} (see also
\cite{Illge}). F.~Andersson described an alternative and considerably
simpler proof of the existence of that potential, both in a spinor
formalism and in tensor formalism
\cite{EdgarHoglund,FAndersson:thesis}. In the new proof an explicit
restriction to four dimensions arises, but the signature can be
arbitrary. In dimensions higher than four one can, using symbolic
computer algebra, derive an integrability condition for the Lanczos
equations which does not appear to be identically satisfied.  It would
be of interest to determine whether this is a real obstruction, or
whether this is simply a new identity which relates the objects at
hand.

M.~Iriondo was supposed to present some results, in collaboration with
E.~Leguizam\'on and O.~Reula, concerning the Newtonian limit of
general relativity on asymptotically null foliations. Unfortunately
she could not attend the conference, and N.~O`Murchadha was kind
enough to give a replacement talk at very short notice. He presented
the results of his joint studies with J.~Guven concerning the
question, ``How Big Can a Spherically Symmetric Static Object Be?''.

It is regrettable that some further speakers could not attend the
meeting. For instance, A.~Rendall was supposed to present his results
in \cite{rendall:2kv,Rendall:wcna} concerning foliations by spacelike
hypersurfaces in space--times with two--dimensional symmetry groups.
He proves in particular that space--times containing such foliations
have crushing singularities for a class of matter models.  J.~Isenberg
was supposed to talk about some new results obtained with A.~Rendall
in \cite{IsenbergRendall}: in this paper they show that there exist
maximal globally hyperbolic solutions of the Einstein--dust equations
which admit a constant mean curvature Cauchy surface, but do not admit
a constant mean curvature foliation.  This is a sharpening of a
previous observation of Rendall \cite{Rendall:wcna}.

\providecommand{\bysame}{\leavevmode\hbox to3em{\hrulefill}\thinspace}

\end{document}